\documentclass[fleqn,usenatbib]{mnras}

\usepackage[T1]{fontenc}

\DeclareRobustCommand{\VAN}[3]{#2}
\let\VANthebibliography\thebibliography
\def\thebibliography{\DeclareRobustCommand{\VAN}[3]{##3}\VANthebibliography}

\usepackage{graphicx}	
\usepackage{amsmath}	
\usepackage{amssymb}	
\usepackage{caption}
\usepackage{subcaption}
\usepackage{ulem}
\usepackage{mathrsfs}
\usepackage{newtxtext,newtxmath}

\title[Optical QPO of AO 0235+164]{Detection of a quasi-periodic oscillation in the optical light curve of 
the remarkable blazar AO 0235+164}

\author[Roy et al. 2022]
{Abhradeep Roy,$^{1}$\thanks{E-mail: \href{mailto:abhradeep.roy@tifr.res.in}{abhradeep.roy@tifr.res.in}}
Varsha R. Chitnis,$^{1}$\thanks{E-mail: \href{mailto:vchitnis@tifr.res.in}{vchitnis@tifr.res.in}}
Alok C. Gupta,$^{2}$\thanks{E-mail: \href{mailto:acgupta30@gmail.com}{acgupta30@gmail.com}}
Paul J. Wiita,$^{3}$
Gustavo E. Romero,$^{4,5}$
\newauthor
Sergio A. Cellone,$^{6,5}$
Anshu Chatterjee,$^{1}$
Jorge A. Combi,$^{4,5,7}$
Claudia M. Raiteri,$^{8}$
Arkadipta Sarkar,$^{1}$
\newauthor
Massimo Villata$^{8}$
\\
\\
$^{1}$Department of High Energy Physics, Tata Institute of Fundamental Research, Homi Bhabha Road, Mumbai-400005, India\\
$^{2}$Aryabhatta Research Institute of Observational Sciences (ARIES), Manora Peak, Nainital, 263001, India \\
$^{3}$Department of Physics, The College of New Jersey, 2000 Pennington Rd., Ewing, NJ 08628-0718, USA\\
$^{4}$Instituto Argentino de Radioastronom\'\i a (CCT-La Plata, CONICET; CICPBA; UNLP), Buenos Aires, Argentina\\
$^{5}$Facultad de Ciencias Astron\'omicas y Geof\'\i sicas, Universidad Nacional de La Plata, La Plata, Buenos Aires, Argentina\\
$^{6}$Complejo Astron\'omico El Leoncito (CASLEO, CONICET-UNLP-UNC-UNSJ), San Juan, Argentina\\
$^{7}$Deptamento de Ingenier\'ia Mec\'anica y Minera, Universidad de Ja\'en, Campus Las Lagunillas s/n Ed. A3 Ja\'en, 23071, Spain \\
$^{8}$INAF-Osservatorio Astrofisico di Torino, Via Osservatorio 20, I-10025 Pino Torinese, Italy 
}

\date{Accepted XXX. Received YYY; in original form ZZZ}

\pubyear{2022}

\begin{document}
\label{firstpage}
\pagerange{\pageref{firstpage}--\pageref{lastpage}}
\maketitle

\begin{abstract}
We present a long term optical $R$ band light curve analysis of the gravitationally lensed blazar AO 0235+164 in the time span 1982 -- 2019. Several methods of analysis lead to the result that there is a periodicity of $\sim$8.13 years present in these data.  In addition,  each of these five major flares are apparently double-peaked, with the secondary peak following the primary one by $\sim$2 years. Along with the well known system, OJ 287, our finding constitutes one of the most secure cases of long term quasi-periodic optical behaviour in a blazar ever found. A binary supermassive black hole system appears to provide a good explanation for these results.
\end{abstract}




\section{Introduction}
\label{sec:intro}

Active galactic nuclei (AGN), presumed to be powered by accreting supermassive black holes (SMBHs) with masses of 10$^{6} -$ 10$^{10}$ M$_{\odot}$, have aspects that can be considered as scaled-up versions of the black hole (BH) and neutron star binaries in the Milky Way and nearby galaxies \citep{1996A&A...308..321F}. The presence of quasi-periodic oscillations (QPOs) in the light curves (LCs) of both the BH and neutron star binaries are fairly common \citep{2006ARA&A..44...49R} but extremely rare in AGN \citep{2018Galax...6....1G}. Among the AGN, blazars (i.e. BL Lac objects (BL Lacs) and flat-spectrum radio quasars (FSRQs)) display the most extreme properties, including violent flux variability across the entire electromagnetic (EM) spectrum, high and variable polarization from radio to optical bands, compact radio structure, the superluminal motion of radio components, and non-thermal emission dominating the whole EM spectrum. These noticeable characteristics of blazars are explained by assuming that the emission comes from a relativistic plasma jet closely aligned with the observer's line of sight \citep[e.g.,][]{1978PhyS...17..265B,1995PASP..107..803U}. \\
\\
AO 0235+164\footnote{\url{https://www.lsw.uni-heidelberg.de/projects/extragalactic/charts/0235+164.html}} ($\alpha_{\rm 2000}$ = 02h 38m 38.9301s; $\delta_{\rm 2000} = +16^{\circ} 36^{\prime} 59.275^{\prime\prime}$) at $z =$ 0.94 \citep{1987ApJ...318..577C} was long ago identified as a BL Lac \citep{1975ApJ...201..275S}. Optical photometric and spectroscopic observations of the source have disclosed foreground absorbing systems at $z =$ 0.524 and $z =$ 0.851 \citep{1987ApJ...318..577C,1996A&A...314..754N}. The flux of the blazar will be contaminated and partially absorbed by the foreground galaxies, the stars of which might act as gravitational micro-lenses. \citet{1988A&A...198L..13S} carried out optical $R$-band photometry, followed by spectroscopic observations with the 2.2 meter and 3.5 meter telescopes, respectively, at Calar Alto, Spain. They reported that AO 0235+164 is unusual with extraordinary properties, apparently arising from that intervening matter.
Hence, this blazar may give insights into the interplay between blazar physics and gravitational microlensing.   \\
\\
Variability of blazars in the whole EM spectrum is essentially stochastic \citep[e.g.,][]{2017ApJ...849..138K}, but occasionally there have been apparent detections of QPOs in the LCs of various blazars in different EM bands on diverse timescales \citep[e.g.,][and references therein]{2008ApJ...679..182E,2009ApJ...690..216G,2014JApA...35..307G,2015Natur.518...74G,Ackermann_2015,2016ApJ...820...20S,2016ApJ...832...47B,2017ApJ...835..260Z,2018NatCo...9.4599Z,2018Galax...6....1G,2019MNRAS.484.5785G,2019MNRAS.487.3990B,2020A&A...642A.129S,2021MNRAS.501...50S,2021MNRAS.501.5997T,2022MNRAS.510.3641R} and several other AGNs \citep[e.g.,][and references therein]{2008Natur.455..369G,2013ApJ...776L..10L,2014JApA...35..307G,2014ApJ...788...31H,2014MNRAS.445L..16A,2015MNRAS.449..467A,2016ApJ...819L..19P,2017ApJ...849....9Z,2018ApJ...853..193Z,2018Galax...6....1G,2018A&A...616L...6G}. AO 0235+164 is one of the blazars in which QPOs have been reported in different EM bands with diverse periods in different epochs of observations. By using optical and radio data of AO 0235+164 taken during 1975 -- 2000, \citet{2001A&A...377..396R} noted that this blazar showed possibly correlated periodic radio and optical outbursts with a period of 5.7$\pm$0.5 years. In a further, longer-term optical data analysis of the source, \citet{2006A&A...459..731R} estimated that the period for the optical outbursts was $\sim$8 years. In a more recent paper, \citet{2017ApJ...837...45F} claimed to have found QPOs with two periods of 8.26 years and 0.54 -- 0.56 years in the historical $R$-band optical data taken during JD 2445300 (26 November 1982) to JD 2456247 (15 November 2012). By analysing {\it All-Sky Monitor (ASM)} X-ray data of the {\it Rossi X-ray Timing Explorer (RXTE)}, \citet{2009ApJ...696.2170R} have reported a possible QPO in AO 0235+164 with a period of $\sim$18 days. In a recent study, \citet{2021MNRAS.501.5997T} reported the detection of a QPO in AO 0235+164 with a dominant period of $\sim$965 days in 4.8, 8.0, and 14.5 GHz radio observations taken over 32 years made at the {\it University of Michigan Radio Astronomy Observatory (UMRAO)}. Hence, the QPOs reported in this source in different EM bands during different epochs have different periods. \\
\\
In 2008, we began a project to search for QPOs in blazars and other subclasses of AGNs in different EM bands on diverse timescales \citep[e.g.,][and references therein]{2009ApJ...690..216G,2018A&A...616L...6G,2009A&A...506L..17L,2009ApJ...696.2170R,2014JApA...35..307G,2018Galax...6....1G,2020MNRAS.499..653K,2020A&A...642A.129S,2021MNRAS.501...50S,2021MNRAS.501.5997T,2022MNRAS.510.3641R}. The detection of QPOs in blazars and other subclasses of AGNs are both rare and rather challenging to explain within existing AGNs models. Here we present a search for  QPOs in the peculiar blazar AO 0235+164 using long term optical data taken in 1982 -- 2019. \\
\\
The paper is structured as follows. In \autoref{sec:obs}, we discuss the new optical observations as well as data from published papers and public archives of the blazar AO 0235+164. In \autoref{sec:ana}, we provide brief explanations of the various analysis techniques we have employed and the obtained results. We present the discussion and conclusions in \autoref{sec:disc} and \autoref{sec:conc}, respectively.

\section{Observations and Public Archive Data}
\label{sec:obs}
\begin{figure*}
    \centering
    \includegraphics[trim={3cm 0 3cm 1cm},clip,width=\textwidth]{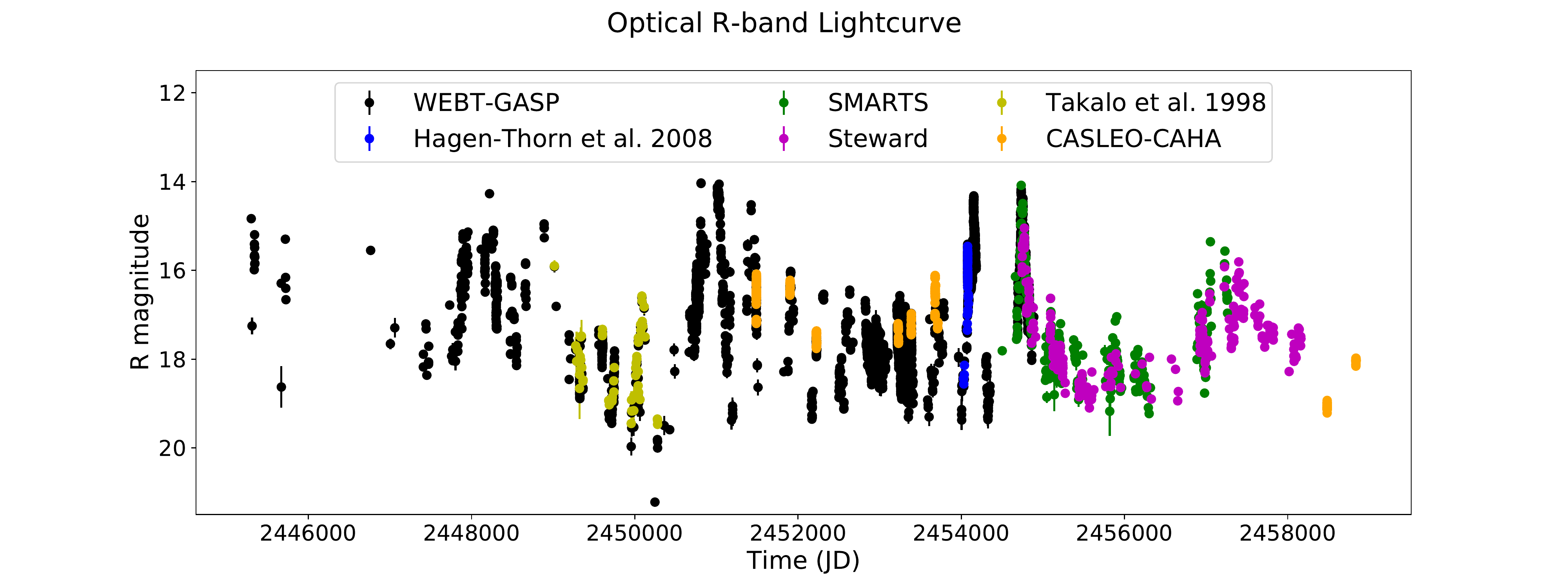}
    \caption{Optical $R$-band light curve of AO 0235+164 measured between JD 2445300 (26 November 1982) and JD 2458835
(17 December 2019).}
    \label{fig:fig1}
\end{figure*}

\begin{figure*}
\centering
\begin{subfigure}{\textwidth}
    \includegraphics[trim={3cm 0 3cm 1.5cm},clip,width=\textwidth]{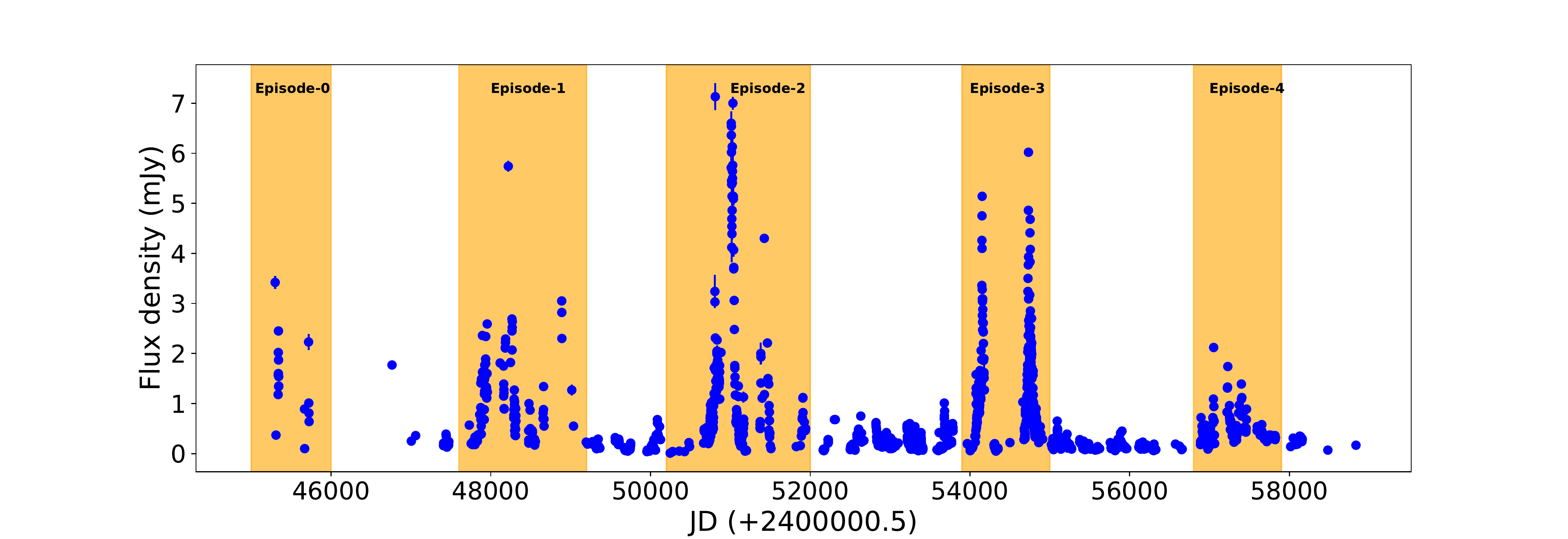}
    \caption{}
    \label{fig:fig2a}
\end{subfigure}

\begin{subfigure}{0.49\textwidth}
    \includegraphics[width=\textwidth]{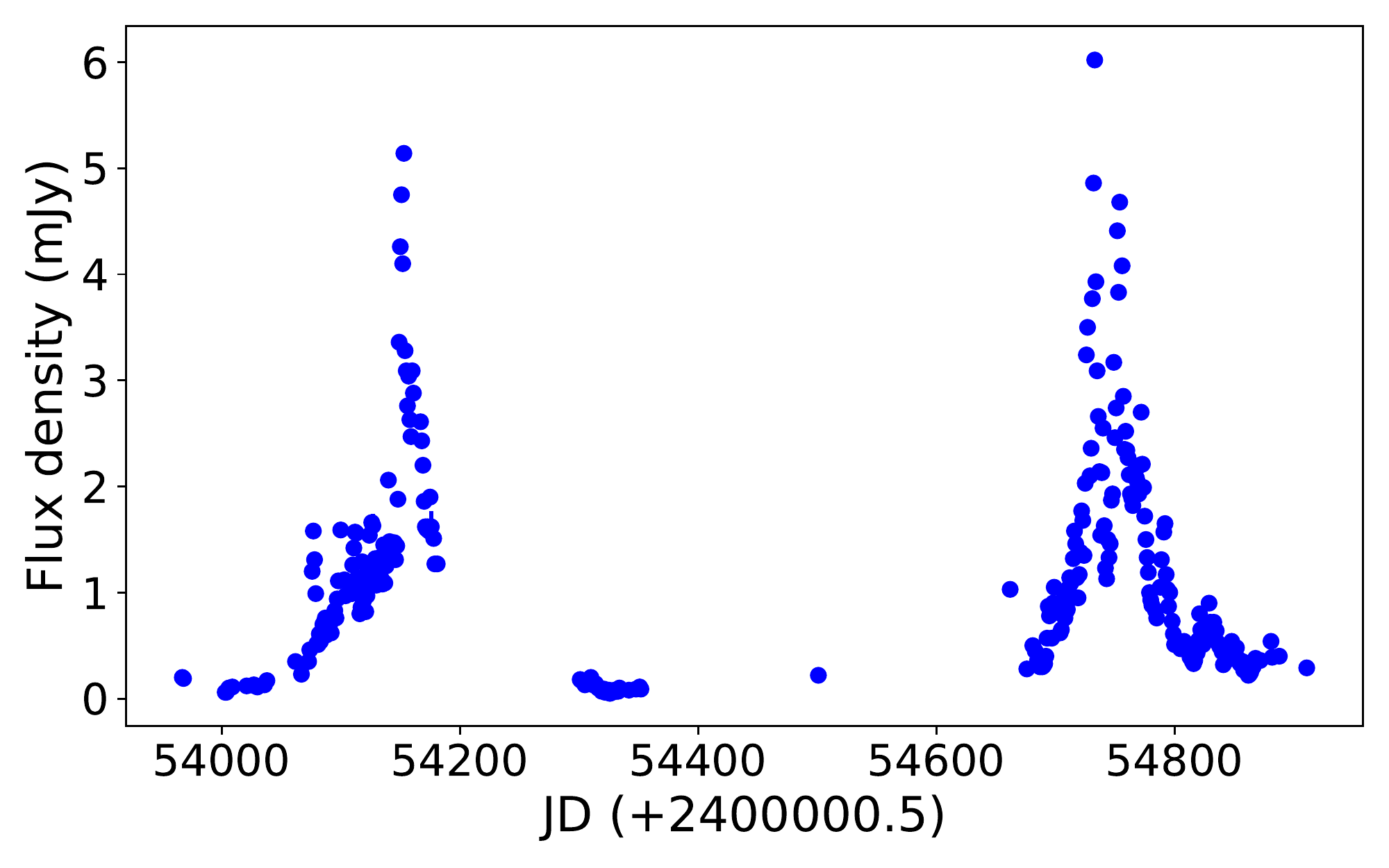}
    \caption{}
    \label{fig:fig2b}
\end{subfigure}
\begin{subfigure}{0.49\textwidth}
    \includegraphics[width=\textwidth]{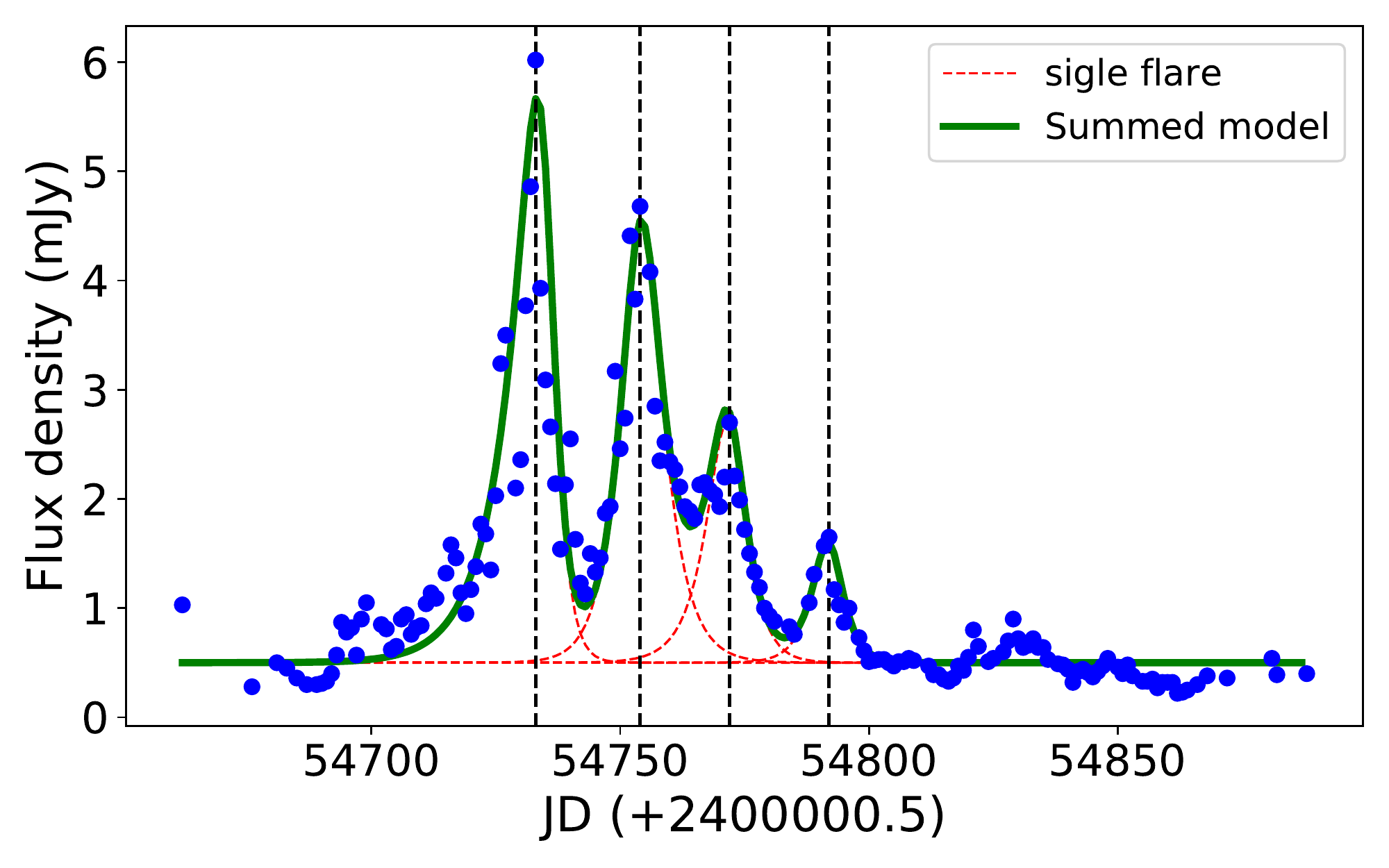}
    \caption{}
    \label{fig:fig2c}
\end{subfigure}
\caption{{\bf(a)} The full optical $R$-band 1-day binned light curve in flux density units (mJy). The five episodes of high activity are highlighted in yellow: Episode-0 (JD 2445000--2446000), Episode-1 (JD 2447600--2449200), Episode-2 (JD 2450200--2452000), Episode-3 (JD 2453900--2455000), and Episode-4 (JD 2456800--2457900). {\bf(b)} Zoomed version of Episode-3. {\bf(c)} Zoomed version of the second flare of Episode-3. All the sub-flares are fitted using \autoref{eq:eq1} (red dashed lines). The green solid line represents the summed model. The fit parameters are given in \autoref{tab:flare_fit}.}
\end{figure*}

The observations of AO 0235+164 newly reported here were carried out in 2018 and 2019 with the 2.15\,m telescope\footnote{\url{https://casleo.conicet.gov.ar/js/}} of Complejo Astron{\'o}mico El Leoncito (CASLEO, Argentina) using a Roper Scientific Versarray 2048B CCD camera (gain: 2.2 e$^{-}$/ADU; read-out noise: 3.1 e$^{-}$). Previous observations with the same telescope (which was equipped with a Tektronicx TK-1024 CCD [gain:1.98 e$^{-}$/ADU; read-out noise: 9.8 e$^{-}$] in 2001 and before) are also used and were originally published in \citet{RCC00aa, RCCA02, 2005A&A...438...39R} and \citet{2006A&A...459..731R}. We also include data taken at the 2.2\,m telescope\footnote{\url{http://www.caha.es/CAHA/Telescopes/2.2m.html}} of Calar Alto Astronomical Observatory (CAHA, Spain) which were published in \citet{CRCM07}. CAHA data were taken in 2005 with the CAFOS instrument in imaging polarimetry mode, with a SITE\#1d CCD (gain: 2.3 electrons/adu; read-out noise: 5.06 electrons). In this case, each polarimetric point was computed from a set of four images taken with a half-wave plate retarder at different angles ($0^\circ$, $22.5^\circ$, $45.0^\circ$, $67.5^\circ$), while photometric data were obtained by adding up the ordinary and extraordinary fluxes from each image. Image processing was done using the \textsc{iraf} package, particularly employing \textsc{apphot} for aperture photometry. \\ 
\\
The largest set of observations used in this work is from The Whole Earth Blazar Telescope\footnote{\url{https://www.oato.inaf.it/blazars/webt}} (WEBT) \citep{2002A&A...390..407V,2017Natur.552..374R} which is an international collaboration of optical, near-infrared, and radio observers. WEBT was initiated in 1997 and since then has organised tens of monitoring campaigns on selected blazars, with the participation of many tens of observers and telescopes all around the world. AO 0235+164 is one of the sources studied by the WEBT and by its GLAST-AGILE Support Program (GASP) \citep{2008A&A...481L..79V,2009A&A...504L...9V}, which was started in 2007 in view of the launch of the {\it AGILE} and {\it Fermi} (formerly {\it GLAST}) satellites. WEBT/GASP data on AO 0235+164 were published in \citet{2001A&A...377..396R,2005A&A...438...39R,2008A&A...480..339R} and \citet{2012ApJ...751..159A}. In particular, \citet{2005A&A...438...39R} gave prescriptions to subtract the contribution of the southern galaxy (called ELISA) from the optical flux densities. These authors also estimated the amount of absorption towards the source in excess of that from our Galaxy in  X-ray, ultraviolet, optical, and near-infrared bands. \\
\\
The data collected during both the WEBT campaigns and the GASP continuous monitoring efforts are calibrated according to common prescriptions, i.e., with the same photometry for the same reference stars. In the case of AO 0235+164, the adopted photometric sequence includes stars 1, 2, and 3 from \citet{1985AJ.....90.1184S}. Minor systematic offsets between the various datasets can nonetheless still be present and are corrected for when building the light curves. Moreover, binning and removal of clear outliers are performed to obtain light curves as reliable as possible for a meaningful further data analysis. \\
\\
In addition, we used the public archive\footnote{\url{http://james.as.arizona.edu/~psmith/Fermi/DATA/Rphotdata.html}} optical $R$-band photometric observations of the blazar AO 0235+164 that were carried out by P.~Smith at Steward Observatory, University of Arizona, using the 2.3 m Bok and 1.54 m Kuiper telescopes. These photometric observations were carried out between 4 October 2008 and 12 February 2018 using the SPOL CCD Imaging/Spectropolarimeter attached to those two telescopes. Details about the instrument, observation, and data analysis are given in detail in \citet{2009arXiv0912.3621S}. We also incorporated the public archive\footnote{\url{http://www.astro.yale.edu/smarts/glast/home.php\#}} optical $R$-band photometric data of the blazar AO 0235+164 from {\it Small and Moderate Aperture Research Telescope System (SMARTS)} telescopes. The SMARTS consortium is part of the Cerro Tololo InterAmerican Observatory (CTIO), Chile, and has been observing Fermi-Large Area Telescope (LAT)-monitored blazars in the optical $B$, $V$, $R$ and NIR $J$ and $K$ bands. Details about the SMARTS telescopes, detectors, observations, and data analysis are provided in \citet{2012ApJ...756...13B}. Besides these two public archives, we also included other $R$-band optical photometric data \citep{1998A&AS..129..577T,2001A&A...377..396R,2005A&A...438...39R,2006A&A...459..731R,2008A&A...480..339R,2008ApJ...672...40H}. The totality of the reduced observations we have considered produces the LC shown in \autoref{fig:fig1}. \\
\\
\autoref{fig:fig2a} shows the full optical $R$-band daily-binned LC with magnitudes converted to flux densities (mJy). We can specify five epochs when AO 0235+164 was in a high flux state: Episode-0 (JD 2445000--2446000), Episode-1 (JD 2447600--2449200), Episode-2 (JD 2450200--2452000), Episode-3 (JD 2453900--2455000), and Episode-4 (JD 2456800--2457900). Although sampling differs between these five high-flux epochs, a general behaviour showing two distinct flares can be appreciated in each of them. Episode-3 has the best data sampling, so it is possible to look for any flare substructure. \autoref{fig:fig2b} zooms into Episode-3. The decaying part of the first flare is not well sampled. However, the second flare is very well sampled and shows four clear flaring substructures with monotonically decreasing amplitudes. We considered a base flux density of 0.5 mJy from the low flux region and fitted each sub-flare by adding to it a time-dependent asymmetric flux profile given as,
\begin{equation}
\label{eq:eq1}
    F(t) = 2 F_0 [\text{e}^{(T_0-t)/T_\mathrm{R}}+\text{e}^{(t-T_0)/T_\mathrm{D}}]^{-1}
\end{equation}
where $F(t)$ is the flux at time $t$, $F_0$ is the flux at time $T_0$, while $T_\mathrm{R}$ and $T_\mathrm{D}$ are respectively the rise and decay timescales of the sub-flare. The parameters of these fits are given in \autoref{tab:flare_fit}.

\begin{table}
\caption{Subflare fitting parameters of the second flare of Episode-3.}
    \centering
    \begin{tabular}{ccccc}
        \hline\hline
        Sub-flare & $T_0$ & $T_R$ & $T_D$ & $F_0$ \\
                 & (JD)  & (days) & (days) & (mJy) \\
        \hline
        1 & 2454735 & 6.2 & 2.0 & 4.5 \\
        2 & 2454754 & 3.5 & 4.0 & 4.0 \\
        3 & 2454772 & 4.0 & 3.0 & 2.2 \\
        4 & 2454792 & 3.0 & 2.5 & 1.1 \\
        \hline
    \end{tabular}
    \label{tab:flare_fit}
\end{table}

\section{Analysis Methods and Results}
\label{sec:ana}

\begin{figure*}
\centering
\begin{subfigure}{0.49\textwidth}
\centering
    \includegraphics[width=\textwidth]{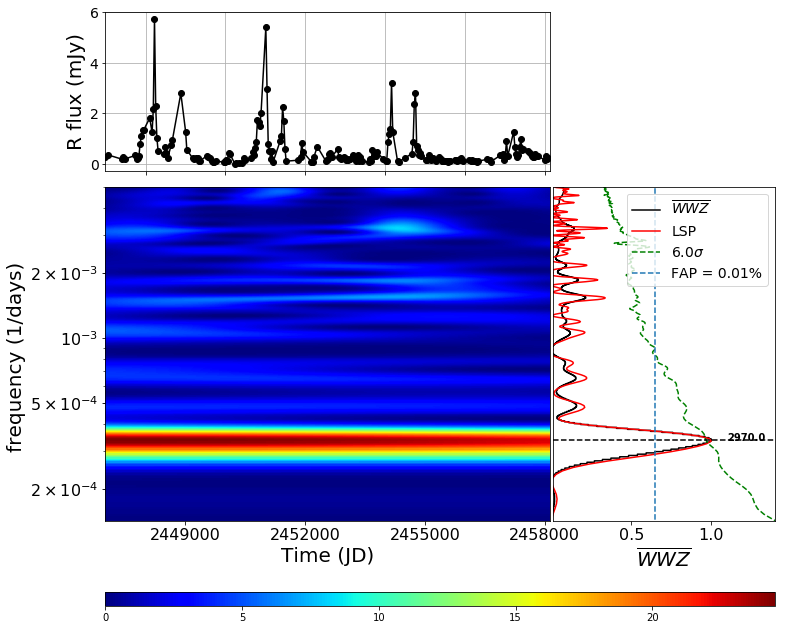}
    \caption{}
    \label{fig:fig3a}    
\end{subfigure}
\begin{subfigure}{0.49\textwidth}
\centering
    \includegraphics[width=\textwidth]{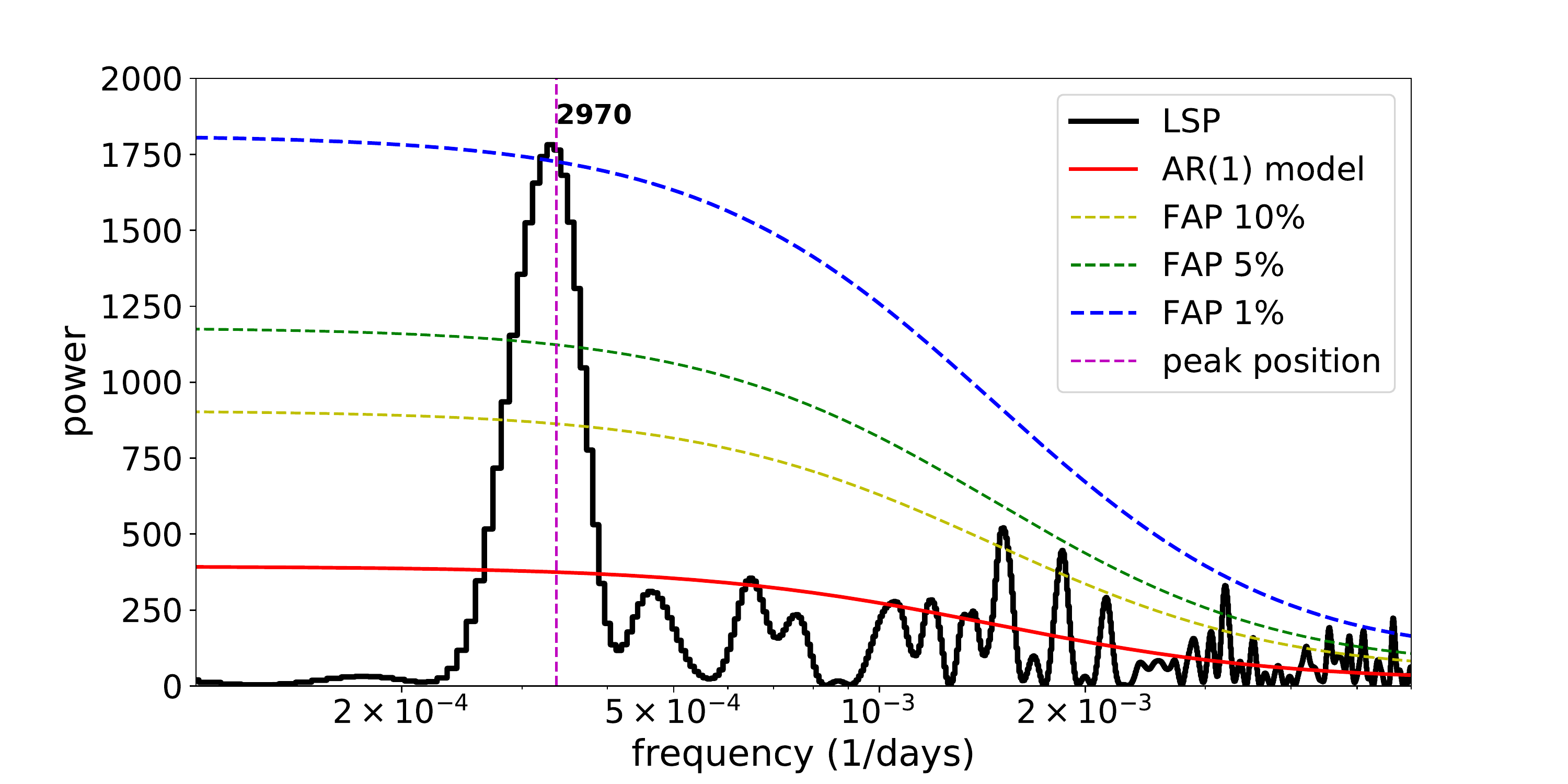}
    \caption{}
    \label{fig:fig3b}    
\end{subfigure}
\caption{\textit{QPO analysis of the AO 0235+164 optical $R$-band long-term 30-day binned light curve:} {\bf (a)} On the upper left sub-panel, the 30-day binned $R$-band light curve (JD 2446994--2458124) is shown with magnitudes converted to flux densities. The lower left sub-panel displays the WWZ map, and the lower right sub-panel contains the time-averaged WWZ power (black line) on top of the normalized GLSP (red line). The dominant peak at 2970 days in the time-averaged WWZ plot crosses the GLSP 0.01 per cent FAP line (light blue dashed line). The green dashed line represents the 6$\sigma$ significance against the bending power-law red-noise spectrum obtained by simulating 2000 light curves. The red band on the WWZ map indicates a strong, $\sim$2970-day (8.13 years) periodicity. {\bf (b)} Power-spectrum of the full $R$-band light curve using REDFIT. The black line represents the power spectrum, the red line is the theoretical AR1 spectrum, and the yellow, green, and blue dashed lines represent the FAP levels of 10, 5, and 1 per cent, respectively. A strong periodicity of $\sim$2970 days crosses the 1 per cent FAP level.}
\end{figure*}

Visual examination of the long-term optical $R$-band light curve in \autoref{fig:fig1} suggests the presence of a persistent QPO. To test for the detection of any periodic modulation and its span in the light curve, and to estimate the corresponding significance, we followed \citet{2022MNRAS.510.3641R} and employed four different methods: Generalized Lomb-Scargle periodogram (GLSP); Weighted Wavelet Z-transform (WWZ); REDFIT; and Monte-Carlo light curve simulations. These methods take independent approaches to detect significant periodicities in an unevenly sampled time series.

\subsection{Generalized Lomb-Scargle periodogram} \label{sec:glsp}

The Lomb-Scargle periodogram (LSP) method fits an unevenly sampled light curve with sinusoids of different frequencies iteratively and produces a periodogram from the goodness of the fit \citep{1976Ap&SS..39..447L, 1982ApJ...263..835S}. The Generalized LSP (GLSP) process is an improvement on the classical LSP as the GLSP takes into account the errors associated with flux measurements in a light curve and uses a sinusoid plus a constant as the fitting function \citep{2009A&A...496..577Z}. We used the GLSP code from the \textsc{pyastronomy} python package\footnote{\url{https://github.com/sczesla/PyAstronomy}} \citep{pya}. This code also estimates the significances of the periodogram peaks in terms of the False Alarm Probability (FAP), considering an underlying white-noise periodogram. \\
\\
In this work, we consider a periodogram peak to be significant if it crosses the 1 per cent FAP line. The periodogram shows the highest power at the 2970-day periodicity (red line in the right sub-panel of \autoref{fig:fig3a}). We fitted a Gaussian to the dominant peak in the periodogram and estimated its position and the corresponding uncertainty to be 2996$^{+75}_{-72}$ days. GLSP cannot effectively detect transient periodicities in a light curve, as the aperiodic part of the light curve reduces the goodness of the sinusoid fit; however, it is an efficient tool to find persistent periodicities.

\subsection{Weighted wavelet Z-transform} \label{sec:wwz}

The weighted wavelet Z-transform (WWZ) method is an important tool for studying any periodicities that may be transient. This method can detect the power of any dominant periodic modulation and its corresponding time span in the light curve. This method creates a WWZ map by convolving the light curve with a time and frequency-dependent kernel and then decomposing the data into time and frequency domains. We convolved the light curve with the Morlet kernel \citep{doi:10.1137/0515056} given as
\begin{equation}
    f[\omega(t-\tau)] = \exp[i\omega(t-\tau) - c\omega^2(t-\tau)^2] .
\end{equation}
Then the WWZ map is given as
\begin{equation}
    W[\omega,\tau; x(t)] = \omega^{1/2} \int x(t)f^{\ast}[\omega(t-\tau)]\text{d}t ,
\end{equation}
where $x(t)$ is the light curve, $f^{\ast}$ is the complex conjugate of the Morlet kernel $f$, $\omega$ is the frequency, and $\tau$ is the time-shift parameter. Such a kernel behaves as a windowed discrete Fourier transform, having a frequency-dependent window size of $\exp[- c\omega^2(t-\tau)^2]$.  \autoref{fig:fig3a} shows the WWZ map of the full $R$-band light curve. The bright red patch on the map indicates a persistent $\sim$2970-day period, and the time-averaged normalized WWZ (black line in the right sub-panel of \autoref{fig:fig3a}) is consistent with the GLSP result.

\subsection{Light curve simulation} \label{sec:lsim}

To estimate the significance of the dominant peak at $\sim$2970 days we simulated 2000 light curves using a Monte Carlo method, where they were generated from the power spectral density (PSD) and the flux distribution (PDF) of the original $R$-band light curve \citep{2013MNRAS.433..907E}. For estimating the true underlying red-noise PSD of our unevenly sampled light curve, we followed \citet{Goyal_2017}. We used a power law to approximate the underlying red-noise PSD and fitted a log-normal model to the PDF of the original light curve. We used the \textsc{delightcurvesimulation}\footnote{\url{https://github.com/samconnolly/DELightcurveSimulation}} code \citep{2016ascl.soft02012C} to simulate the light curves. The mean and standard deviation of the GLSPs at each frequency when calculated for each simulated light curve allowed us to estimate a significance of $\sim$6$\sigma$ for the QPO of $\sim$2970 days.

\subsection{REDFIT} \label{sec:rf}

The \textsc{redfit}\footnote{\url{https://www.manfredmudelsee.com/soft/redfit/index.htm}} software is a FORTRAN 90 program that fits a first-order autoregressive (AR1) process, to an unevenly spaced time series and has the advantage of avoiding any interpolation in the time domain \citep{SCHULZ2002421}. The fitted AR1 model is then transformed from the time domain into the frequency domain to estimate the underlying red-noise spectrum, which is characteristic of the variability of most blazar emissions. Comparison of the spectrum of the actual time series with the estimated red-noise spectrum allows testing the hypothesis that the origin of the light curve is an AR1 process. Critically, this code can be used to test if peaks in the spectrum of a time series are significant against the red-noise background from an AR1 process. \\
\\
\textsc{redfit} estimates the significance of the peaks in the power-spectrum using four FAP levels: 20, 10, 5 and 1 per cent. The \textsc{redfit} periodogram of $R$-band light curve (\autoref{fig:fig3b}) also shows a dominant periodicity at $\sim$2970 days that crosses the 1 per cent FAP level.

\section{discussion}
\label{sec:disc}

In this work, we applied LSP, WWZ, and \textsc{redfit} methods to the very long term optical $R$-band light curve of AO 0235+164, and detected a QPO of $\sim$2970 days (8.13 years) with $\sim$6$\sigma$ significance. Each of the flares is followed by a second peak, with a gap between the double peaks ranging from 1.5--2 years. In the case of BL Lac objects, such as AO 0235+164, the synchrotron radiation from the jet dominates the emission in optical wavebands \citep[e.g.,][]{2006A&A...459..731R,2012ApJ...751..159A}. The high energy $\gamma$-ray emission usually arises from the inverse Compton scattering of the seed photon field by the synchrotron emitting electron population. So we expect to detect a similar QPO in $\gamma$-ray observations. Unfortunately, $\gamma$-ray data for this source are available only for the last 12 years. But still, we can gain a few insights about the underlying physical processes from the optical QPO period and its persistence for at least the five cycles these observations cover. \\
\\
Somewhat similar double-peaked periodic ($\sim$12 years) optical flares, now spanning 10 cycles, have been detected in the blazar OJ 287 \citep[e.g.,][]{2008Natur.452..851V,2019Univ....5..108D,2019ApJ...882...88V,2021Galax..10....1V}. 
The original binary SMBH model for OJ 287 \citep[e.g.,][]{1988ApJ...325..628S} could explain a single large flare per binary period, but not two. The changing intervals between the first  and second flares, together with their flat spectra and non-polarized nature, strongly indicate that in OJ 287 we are seeing  thermal flares arising from the two interactions of the secondary SMBH with the accretion disc of the primary SMBH per orbit \citep{1997ApJ...484..180S,2019ApJ...882...88V,2021Galax..10....1V}. \\
\\
Although we have much less information on AO 0235+164 than on OJ 287, the Fermi-LAT $\gamma$-ray light curve of AO 0235+164 shows emission correlated with optical light curves during its 2008--2009 outburst and the moderately high state in 2014--2016 \citep{2020ApJ...902...41W}. Long term radio light curves also showed correlated variation with the optical $R$-band \citep{2001A&A...377..396R}. Such correlations suggest that the pairs of quasi-periodic flares originate from a jet, not from the accretion disc, while it is unlikely that thermal flares as bright as these would be seen from such a distant source.   \citet{universe7080261} reported correlated two-year-long radio and optical light curves of OJ 287 and described the flares as precursor flares due to temporary accretion and jet launching from the secondary BH during the impact with the accretion disc. A similar process involving the emergence of a secondary jet may be responsible for the paired multi-waveband outbursts in AO 0235+164, assuming a highly eccentric elliptic orbit of the secondary BH. \\
\\
\\
Other models that can explain QPOs in blazars involve changes in the orientation of the emitting regions of an inhomogeneous curved jet \citep{2017Natur.552..374R}. This class of model has been applied to A0~0235+164 by \citet{2004A&A...419..913O}, where the quasi-periodic radio and optical light curves were explained in terms of a rotating helical jet. In this geometrical scenario, the double-peaked structure of the outbursts could be due to the precession of the two jets emerging from the vicinities of the black holes in a binary system, as proposed to describe the centenary optical light curve of OJ~287 by \citet{1998MNRAS.293L..13V}. \\
\\
To explain the subflare structure in the second flare of Episode-3 (\autoref{fig:fig2b}) showing a nearly periodic pattern with monotonically decreasing amplitudes, we may combine the two models discussed previously \citep{1988ApJ...325..628S, 2017Natur.552..374R, 2019Univ....5..108D}. Assuming a precessing highly eccentric elliptical orbit of the secondary BH, the double peaks in a flaring episode may occur due to the injection of magnetized plasma into the jet, as the high amplitude tidal disruptions, caused by the two quick impacts ($\sim$1--2 years apart) of the secondary BH onto the accretion disc, reach the jet base. The resulting ejected plasma disturbance, moving helically down the jet, emits less as it radiates, thus produces the observed subflare structure.\\
\\
Another possible explanation of the double-peaked structure in the optical light curve flares could be the delay caused by gravitational lensing of AO 0235+164, given the presence of absorbing systems at lower redshift along our line of sight. However, deep observations that resolve the foreground lens galaxy at $z=0.524$ also ruled out  
the existence of images at angular distances greater than 0.5 arcsec from the central source \cite[see, e.g.,][]{1993ApJ...415..101A}. This makes it very unlikely that the double peaks with separations of $\sim$2 years in the light curves reported in this paper could be caused by gravitational time delays in the light travel of photons produced in the  same intrinsic event. Additionally, as can be seen from \autoref{fig:fig2c}, the second peak has a quite different substructure from the first peak when observed with higher resolution. This is not compatible with the lensed scenario, where both peaks are different images of the same intrinsic variation.\\ 
\\


\section{Conclusions}
\label{sec:conc}
In the present work, we have presented the longest $R$-band optical light curve observations of the blazar AO 0235+164. The data are taken from the public archives and published papers, as well as from new observations. The data presented here were taken between 26 November 1982 and 17 December 2019, a duration of over 37 years. We searched for QPOs in the light curve using 30-day binned data and found a QPO of $\sim$8.13 years at a high confidence level according to multiple techniques, despite there only being 5 cycles during this period. 
This result is consistent with the period previously found by \citet{2006A&A...459..731R} and \citet{2017ApJ...837...45F}. Even though some of the data in these two papers are common with our study, our analysis, covering a longer time span, strongly confirms the $\sim 8.13$-year QPO.
The most striking feature newly noticed here in this extended light curve is that the main peaks, separated by 8.13 years, are followed by secondary peaks after $\sim$2 years. \\ 
\\
The totality of these observations seems to be understandable if AO 0235+164 contains a binary supermassive black hole system of period $\sim$8 years. More than one variation of this basic scenario could work, as long as the observed emission arises from the jet.  Passage of the secondary through the disc of the primary can trigger a region of enhanced emission moving out through the relativistic jet.  This region could have a helical component to its motion, and its second passage closest to our line-of-sight could provide the second flares followed by $\sim$2 years.  Alternatively, if the orbit is highly elliptical, with the two passages of the secondary through the disc separated by $\sim$2 years, then each such impact could drive an independent jet brightening.\\  
\\
We expect the next such $\sim$2-year long flaring episode to happen between November 2022 and May 2025 (JD 2459900 -- 2460800). An intensive multiwavelength WEBT campaign will be conducted during that period to test the persistence of this apparent QPO in AO~0235+164 and possibly discriminate between these hypotheses.

\section*{Acknowledgements}

Data from the Steward Observatory spectropolarimetric monitoring project were used. This program is supported by Fermi Guest Investigator grants NNX08AW56G, NNX09AU10G, NNX12AO93G, and NNX15AU81G. This paper has made use of up-to-date SMARTS optical/near-infrared light curves that are available at {\url{www.astro.yale.edu/smarts/glast/home.php}}. This work is partly based on data taken and assembled by the WEBT collaboration and stored in the WEBT archive at the Osservatorio Astrofisico di Torino - INAF (\url{https://www.oato.inaf.it/blazars/webt/}). This work is based on data acquired at Complejo Astron\'omico El Leoncito, operated under agreement between the Consejo Nacional de Investigaciones Cient\'ificas y T\'ecnicas de la Rep\'ublica Argentina and the National Universities of La Plata, C\'ordoba and San Juan. \\
\\
We thank the anonymous reviewer for useful suggestions. We thank Anabella Araudo and Ileana Andruchow for help with the observations made with CASLEO and the data analysis. VRC and ACG thank A.\ Gopakumar for discussion. We acknowledge the support of the Department of Atomic Energy, Government of India, under project identification number RTI 4002. JAC is Mar\'ia Zambrano researcher supported by PIP 0113 (CONICET) and PICT-2017-2865 (ANPCyT). JAC was also supported by grant PID2019-105510GB-C32/AEI/10.13039/501100011033 from the Agencia Estatal de Investigaci\'on of the Spanish Ministerio de Ciencia, Innovaci\'on y Universidades, and by Consejer\'{\i}a de Econom\'{\i}a, Innovaci\'on, Ciencia y Empleo of Junta de Andaluc\'{\i}a as research group FQM- 322, as well as FEDER funds.

\section*{Data Availability}
    Data acquired by the WEBT collaboration are stored in the WEBT archive and are available upon request to the WEBT President Massimo Villata (\href{mailto:massimo.villata@inaf.it}{massimo.villata@inaf.it}).
    Data from Steward Observatory are publicly available at \url{http://james.as.arizona.edu/~psmith/Fermi/DATA/Rphotdata.html}. The data collected by the SMARTS telescope are available publicly at \url{http://www.astro.yale.edu/smarts/glast/home.php\#}. The VizieR $BVRI$ photometry data published in \citet{1998A&AS..129..577T} are available at \url{https://cdsarc.cds.unistra.fr/viz-bin/cat/J/A+AS/129/577#/browse}.



\bibliographystyle{mnras}
\bibliography{references} 

\begin{thebibliography}{}
\makeatletter
\relax
\def\mn@urlcharsother{\let\do\@makeother \do\$\do\&\do\#\do\^\do\_\do\%\do\~}
\def\mn@doi{\begingroup\mn@urlcharsother \@ifnextchar [ {\mn@doi@}
  {\mn@doi@[]}}
\def\mn@doi@[#1]#2{\def\@tempa{#1}\ifx\@tempa\@empty \href
  {http://dx.doi.org/#2} {doi:#2}\else \href {http://dx.doi.org/#2} {#1}\fi
  \endgroup}
\def\mn@eprint#1#2{\mn@eprint@#1:#2::\@nil}
\def\mn@eprint@arXiv#1{\href {http://arxiv.org/abs/#1} {{\tt arXiv:#1}}}
\def\mn@eprint@dblp#1{\href {http://dblp.uni-trier.de/rec/bibtex/#1.xml}
  {dblp:#1}}
\def\mn@eprint@#1:#2:#3:#4\@nil{\def\@tempa {#1}\def\@tempb {#2}\def\@tempc
  {#3}\ifx \@tempc \@empty \let \@tempc \@tempb \let \@tempb \@tempa \fi \ifx
  \@tempb \@empty \def\@tempb {arXiv}\fi \@ifundefined
  {mn@eprint@\@tempb}{\@tempb:\@tempc}{\expandafter \expandafter \csname
  mn@eprint@\@tempb\endcsname \expandafter{\@tempc}}}

\bibitem[\protect\citeauthoryear{{Abraham}, {Crawford}, {Merrifield},
  {Hutchings}  \& {McHardy}}{{Abraham} et~al.}{1993}]{1993ApJ...415..101A}
{Abraham} R.~G.,  {Crawford} C.~S.,  {Merrifield} M.~R.,  {Hutchings} J.~B.,
  {McHardy} I.~M.,  1993, \mn@doi [\apj] {10.1086/173147}, \href
  {https://ui.adsabs.harvard.edu/abs/1993ApJ...415..101A} {415, 101}

\bibitem[\protect\citeauthoryear{{Ackermann} et~al.,}{{Ackermann}
  et~al.}{2012}]{2012ApJ...751..159A}
{Ackermann} M.,  et~al., 2012, \mn@doi [\apj] {10.1088/0004-637X/751/2/159},
  \href {https://ui.adsabs.harvard.edu/abs/2012ApJ...751..159A} {751, 159}

\bibitem[\protect\citeauthoryear{Ackermann et~al.,}{Ackermann
  et~al.}{2015}]{Ackermann_2015}
Ackermann M.,  et~al., 2015, \mn@doi [\apj] {10.1088/2041-8205/813/2/l41}, 813,
  L41

\bibitem[\protect\citeauthoryear{{Alston}, {Markeviciute}, {Kara}, {Fabian}  \&
  {Middleton}}{{Alston} et~al.}{2014}]{2014MNRAS.445L..16A}
{Alston} W.~N.,  {Markeviciute} J.,  {Kara} E.,  {Fabian} A.~C.,   {Middleton}
  M.,  2014, \mn@doi [\mnras] {10.1093/mnrasl/slu127}, \href
  {https://ui.adsabs.harvard.edu/abs/2014MNRAS.445L..16A} {445, L16}

\bibitem[\protect\citeauthoryear{{Alston}, {Parker},
  {Markevi{\v{c}}i{\={u}}t{\.{e}}}, {Fabian}, {Middleton}, {Lohfink}, {Kara}
  \& {Pinto}}{{Alston} et~al.}{2015}]{2015MNRAS.449..467A}
{Alston} W.~N.,  {Parker} M.~L.,  {Markevi{\v{c}}i{\={u}}t{\.{e}}} J.,
  {Fabian} A.~C.,  {Middleton} M.,  {Lohfink} A.,  {Kara} E.,   {Pinto} C.,
  2015, \mn@doi [\mnras] {10.1093/mnras/stv351}, \href
  {https://ui.adsabs.harvard.edu/abs/2015MNRAS.449..467A} {449, 467}

\bibitem[\protect\citeauthoryear{{Bhatta}}{{Bhatta}}{2019}]{2019MNRAS.487.3990B}
{Bhatta} G.,  2019, \mn@doi [\mnras] {10.1093/mnras/stz1482}, \href
  {https://ui.adsabs.harvard.edu/abs/2019MNRAS.487.3990B} {487, 3990}

\bibitem[\protect\citeauthoryear{{Bhatta} et~al.,}{{Bhatta}
  et~al.}{2016}]{2016ApJ...832...47B}
{Bhatta} G.,  et~al., 2016, \mn@doi [\apj] {10.3847/0004-637X/832/1/47}, \href
  {https://ui.adsabs.harvard.edu/abs/2016ApJ...832...47B} {832, 47}

\bibitem[\protect\citeauthoryear{{Blandford} \& {Rees}}{{Blandford} \&
  {Rees}}{1978}]{1978PhyS...17..265B}
{Blandford} R.~D.,  {Rees} M.~J.,  1978, \mn@doi [\physscr]
  {10.1088/0031-8949/17/3/020}, \href
  {https://ui.adsabs.harvard.edu/abs/1978PhyS...17..265B} {17, 265}

\bibitem[\protect\citeauthoryear{{Bonning} et~al.,}{{Bonning}
  et~al.}{2012}]{2012ApJ...756...13B}
{Bonning} E.,  et~al., 2012, \mn@doi [\apj] {10.1088/0004-637X/756/1/13}, \href
  {https://ui.adsabs.harvard.edu/abs/2012ApJ...756...13B} {756, 13}

\bibitem[\protect\citeauthoryear{{Cellone}, {Romero}, {Combi}  \&
  {Mart\'{\i}}}{{Cellone} et~al.}{2007}]{CRCM07}
{Cellone} S.~A.,  {Romero} G.~E.,  {Combi} J.~A.,   {Mart\'{\i}} J.,  2007,
  \mn@doi [\mnras] {10.1111/j.1745-3933.2007.00366.x}, 381, L60

\bibitem[\protect\citeauthoryear{{Cohen}, {Smith}, {Junkkarinen}  \&
  {Burbidge}}{{Cohen} et~al.}{1987}]{1987ApJ...318..577C}
{Cohen} R.~D.,  {Smith} H.~E.,  {Junkkarinen} V.~T.,   {Burbidge} E.~M.,  1987,
  \mn@doi [\apj] {10.1086/165393}, \href
  {https://ui.adsabs.harvard.edu/abs/1987ApJ...318..577C} {318, 577}

\bibitem[\protect\citeauthoryear{{Connolly}}{{Connolly}}{2016}]{2016ascl.soft02012C}
{Connolly} S.~D.,  2016, {DELightcurveSimulation: Light curve simulation code}
  (\mn@eprint {ascl} {1602.012})

\bibitem[\protect\citeauthoryear{{Czesla}, {Schr{\"o}ter}, {Schneider},
  {Huber}, {Pfeifer}, {Andreasen}  \& {Zechmeister}}{{Czesla}
  et~al.}{2019}]{pya}
{Czesla} S.,  {Schr{\"o}ter} S.,  {Schneider} C.~P.,  {Huber} K.~F.,  {Pfeifer}
  F.,  {Andreasen} D.~T.,   {Zechmeister} M.,  2019, {PyA: Python
  astronomy-related packages} (\mn@eprint {ascl} {1906.010})

\bibitem[\protect\citeauthoryear{{Dey} et~al.,}{{Dey}
  et~al.}{2019}]{2019Univ....5..108D}
{Dey} L.,  et~al., 2019, \mn@doi [Universe] {10.3390/universe5050108}, \href
  {https://ui.adsabs.harvard.edu/abs/2019Univ....5..108D} {5, 108}

\bibitem[\protect\citeauthoryear{{Emmanoulopoulos}, {McHardy}  \&
  {Papadakis}}{{Emmanoulopoulos} et~al.}{2013}]{2013MNRAS.433..907E}
{Emmanoulopoulos} D.,  {McHardy} I.~M.,   {Papadakis} I.~E.,  2013, \mn@doi
  [\mnras] {10.1093/mnras/stt764}, \href
  {https://ui.adsabs.harvard.edu/abs/2013MNRAS.433..907E} {433, 907}

\bibitem[\protect\citeauthoryear{{Espaillat}, {Bregman}, {Hughes}  \&
  {Lloyd-Davies}}{{Espaillat} et~al.}{2008}]{2008ApJ...679..182E}
{Espaillat} C.,  {Bregman} J.,  {Hughes} P.,   {Lloyd-Davies} E.,  2008,
  \mn@doi [\apj] {10.1086/587023}, \href
  {https://ui.adsabs.harvard.edu/abs/2008ApJ...679..182E} {679, 182}

\bibitem[\protect\citeauthoryear{{Falcke} \& {Biermann}}{{Falcke} \&
  {Biermann}}{1996}]{1996A&A...308..321F}
{Falcke} H.,  {Biermann} P.~L.,  1996, \aap, \href
  {https://ui.adsabs.harvard.edu/abs/1996A&A...308..321F} {308, 321}

\bibitem[\protect\citeauthoryear{{Fan} et~al.,}{{Fan}
  et~al.}{2017}]{2017ApJ...837...45F}
{Fan} J.~H.,  et~al., 2017, \mn@doi [\apj] {10.3847/1538-4357/aa5def}, \href
  {https://ui.adsabs.harvard.edu/abs/2017ApJ...837...45F} {837, 45}

\bibitem[\protect\citeauthoryear{{Gierli{\'n}ski}, {Middleton}, {Ward}  \&
  {Done}}{{Gierli{\'n}ski} et~al.}{2008}]{2008Natur.455..369G}
{Gierli{\'n}ski} M.,  {Middleton} M.,  {Ward} M.,   {Done} C.,  2008, \mn@doi
  [\nat] {10.1038/nature07277}, \href
  {https://ui.adsabs.harvard.edu/abs/2008Natur.455..369G} {455, 369}

\bibitem[\protect\citeauthoryear{Goyal et~al.,}{Goyal
  et~al.}{2017}]{Goyal_2017}
Goyal A.,  et~al., 2017, \mn@doi [The Astrophysical Journal]
  {10.3847/1538-4357/aa6000}, 837, 127

\bibitem[\protect\citeauthoryear{{Graham} et~al.,}{{Graham}
  et~al.}{2015}]{2015Natur.518...74G}
{Graham} M.~J.,  et~al., 2015, \mn@doi [\nat] {10.1038/nature14143}, \href
  {https://ui.adsabs.harvard.edu/abs/2015Natur.518...74G} {518, 74}

\bibitem[\protect\citeauthoryear{Grossmann \& Morlet}{Grossmann \&
  Morlet}{1984}]{doi:10.1137/0515056}
Grossmann A.,  Morlet J.,  1984, \mn@doi [SJMA] {10.1137/0515056}, 15, 723

\bibitem[\protect\citeauthoryear{{Gupta}}{{Gupta}}{2014}]{2014JApA...35..307G}
{Gupta} A.~C.,  2014, \mn@doi [Journal of Astrophysics and Astronomy]
  {10.1007/s12036-014-9219-7}, \href
  {https://ui.adsabs.harvard.edu/abs/2014JApA...35..307G} {35, 307}

\bibitem[\protect\citeauthoryear{{Gupta}}{{Gupta}}{2018}]{2018Galax...6....1G}
{Gupta} A.,  2018, \mn@doi [Galaxies] {10.3390/galaxies6010001}, \href
  {https://ui.adsabs.harvard.edu/abs/2018Galax...6....1G} {6, 1}

\bibitem[\protect\citeauthoryear{{Gupta}, {Srivastava}  \& {Wiita}}{{Gupta}
  et~al.}{2009}]{2009ApJ...690..216G}
{Gupta} A.~C.,  {Srivastava} A.~K.,   {Wiita} P.~J.,  2009, \mn@doi [\apj]
  {10.1088/0004-637X/690/1/216}, \href
  {https://ui.adsabs.harvard.edu/abs/2009ApJ...690..216G} {690, 216}

\bibitem[\protect\citeauthoryear{{Gupta}, {Tripathi}, {Wiita}, {Gu}, {Bambi}
  \& {Ho}}{{Gupta} et~al.}{2018}]{2018A&A...616L...6G}
{Gupta} A.~C.,  {Tripathi} A.,  {Wiita} P.~J.,  {Gu} M.,  {Bambi} C.,   {Ho}
  L.~C.,  2018, \mn@doi [\aap] {10.1051/0004-6361/201833629}, \href
  {https://ui.adsabs.harvard.edu/abs/2018A&A...616L...6G} {616, L6}

\bibitem[\protect\citeauthoryear{{Gupta}, {Tripathi}, {Wiita}, {Kushwaha},
  {Zhang}  \& {Bambi}}{{Gupta} et~al.}{2019}]{2019MNRAS.484.5785G}
{Gupta} A.~C.,  {Tripathi} A.,  {Wiita} P.~J.,  {Kushwaha} P.,  {Zhang} Z.,
  {Bambi} C.,  2019, \mn@doi [\mnras] {10.1093/mnras/stz395}, \href
  {https://ui.adsabs.harvard.edu/abs/2019MNRAS.484.5785G} {484, 5785}

\bibitem[\protect\citeauthoryear{{Hagen-Thorn}, {Larionov}, {Jorstad},
  {Arkharov}, {Hagen-Thorn}, {Efimova}, {Larionova}  \&
  {Marscher}}{{Hagen-Thorn} et~al.}{2008}]{2008ApJ...672...40H}
{Hagen-Thorn} V.~A.,  {Larionov} V.~M.,  {Jorstad} S.~G.,  {Arkharov} A.~A.,
  {Hagen-Thorn} E.~I.,  {Efimova} N.~V.,  {Larionova} L.~V.,   {Marscher}
  A.~P.,  2008, \mn@doi [\apj] {10.1086/523841}, \href
  {https://ui.adsabs.harvard.edu/abs/2008ApJ...672...40H} {672, 40}

\bibitem[\protect\citeauthoryear{{Hu}, {Chou}, {Yang}  \& {Su}}{{Hu}
  et~al.}{2014}]{2014ApJ...788...31H}
{Hu} C.-P.,  {Chou} Y.,  {Yang} T.-C.,   {Su} Y.-H.,  2014, \mn@doi [\apj]
  {10.1088/0004-637X/788/1/31}, \href
  {https://ui.adsabs.harvard.edu/abs/2014ApJ...788...31H} {788, 31}

\bibitem[\protect\citeauthoryear{Komossa et~al.,}{Komossa
  et~al.}{2021}]{universe7080261}
Komossa S.,  et~al., 2021, \mn@doi [Universe] {10.3390/universe7080261}, 7

\bibitem[\protect\citeauthoryear{{Kushwaha}, {Sinha}, {Misra}, {Singh}  \& {de
  Gouveia Dal Pino}}{{Kushwaha} et~al.}{2017}]{2017ApJ...849..138K}
{Kushwaha} P.,  {Sinha} A.,  {Misra} R.,  {Singh} K.~P.,   {de Gouveia Dal
  Pino} E.~M.,  2017, \mn@doi [\apj] {10.3847/1538-4357/aa8ef5}, \href
  {https://ui.adsabs.harvard.edu/abs/2017ApJ...849..138K} {849, 138}

\bibitem[\protect\citeauthoryear{{Kushwaha}, {Sarkar}, {Gupta}, {Tripathi}  \&
  {Wiita}}{{Kushwaha} et~al.}{2020}]{2020MNRAS.499..653K}
{Kushwaha} P.,  {Sarkar} A.,  {Gupta} A.~C.,  {Tripathi} A.,   {Wiita} P.~J.,
  2020, \mn@doi [\mnras] {10.1093/mnras/staa2899}, \href
  {https://ui.adsabs.harvard.edu/abs/2020MNRAS.499..653K} {499, 653}

\bibitem[\protect\citeauthoryear{{Lachowicz}, {Gupta}, {Gaur}  \&
  {Wiita}}{{Lachowicz} et~al.}{2009}]{2009A&A...506L..17L}
{Lachowicz} P.,  {Gupta} A.~C.,  {Gaur} H.,   {Wiita} P.~J.,  2009, \mn@doi
  [\aap] {10.1051/0004-6361/200913161}, \href
  {https://ui.adsabs.harvard.edu/abs/2009A&A...506L..17L} {506, L17}

\bibitem[\protect\citeauthoryear{{Lin}, {Irwin}, {Godet}, {Webb}  \&
  {Barret}}{{Lin} et~al.}{2013}]{2013ApJ...776L..10L}
{Lin} D.,  {Irwin} J.~A.,  {Godet} O.,  {Webb} N.~A.,   {Barret} D.,  2013,
  \mn@doi [\apjl] {10.1088/2041-8205/776/1/L10}, \href
  {https://ui.adsabs.harvard.edu/abs/2013ApJ...776L..10L} {776, L10}

\bibitem[\protect\citeauthoryear{{Lomb}}{{Lomb}}{1976}]{1976Ap&SS..39..447L}
{Lomb} N.~R.,  1976, \mn@doi [\apss] {10.1007/BF00648343}, \href
  {https://ui.adsabs.harvard.edu/abs/1976Ap&SS..39..447L} {39, 447}

\bibitem[\protect\citeauthoryear{{Nilsson}, {Charles}, {Pursimo}, {Takalo},
  {Sillanp\"{a}\"{a}}  \& {Teerikorpi}}{{Nilsson}
  et~al.}{1996}]{1996A&A...314..754N}
{Nilsson} K.,  {Charles} P.~A.,  {Pursimo} T.,  {Takalo} L.~O.,
  {Sillanp\"{a}\"{a}} A.,   {Teerikorpi} P.,  1996, \aap, \href
  {https://ui.adsabs.harvard.edu/abs/1996A&A...314..754N} {314, 754}

\bibitem[\protect\citeauthoryear{{Ostorero}, {Villata}  \&
  {Raiteri}}{{Ostorero} et~al.}{2004}]{2004A&A...419..913O}
{Ostorero} L.,  {Villata} M.,   {Raiteri} C.~M.,  2004, \mn@doi [\aap]
  {10.1051/0004-6361:20035813}, \href
  {https://ui.adsabs.harvard.edu/abs/2004A&A...419..913O} {419, 913}

\bibitem[\protect\citeauthoryear{{Pan}, {Yuan}, {Yao}, {Zhou}, {Liu}, {Zhou}
  \& {Zhang}}{{Pan} et~al.}{2016}]{2016ApJ...819L..19P}
{Pan} H.-W.,  {Yuan} W.,  {Yao} S.,  {Zhou} X.-L.,  {Liu} B.,  {Zhou} H.,
  {Zhang} S.-N.,  2016, \mn@doi [\apjl] {10.3847/2041-8205/819/2/L19}, \href
  {https://ui.adsabs.harvard.edu/abs/2016ApJ...819L..19P} {819, L19}

\bibitem[\protect\citeauthoryear{{Raiteri} et~al.,}{{Raiteri}
  et~al.}{2001}]{2001A&A...377..396R}
{Raiteri} C.~M.,  et~al., 2001, \mn@doi [\aap] {10.1051/0004-6361:20011112},
  \href {https://ui.adsabs.harvard.edu/abs/2001A&A...377..396R} {377, 396}

\bibitem[\protect\citeauthoryear{{Raiteri} et~al.,}{{Raiteri}
  et~al.}{2005}]{2005A&A...438...39R}
{Raiteri} C.~M.,  et~al., 2005, \mn@doi [\aap] {10.1051/0004-6361:20042567},
  \href {https://ui.adsabs.harvard.edu/abs/2005A&A...438...39R} {438, 39}

\bibitem[\protect\citeauthoryear{{Raiteri} et~al.,}{{Raiteri}
  et~al.}{2006}]{2006A&A...459..731R}
{Raiteri} C.~M.,  et~al., 2006, \mn@doi [\aap] {10.1051/0004-6361:20065744},
  \href {https://ui.adsabs.harvard.edu/abs/2006A&A...459..731R} {459, 731}

\bibitem[\protect\citeauthoryear{{Raiteri} et~al.,}{{Raiteri}
  et~al.}{2008}]{2008A&A...480..339R}
{Raiteri} C.~M.,  et~al., 2008, \mn@doi [\aap] {10.1051/0004-6361:20079044},
  \href {https://ui.adsabs.harvard.edu/abs/2008A&A...480..339R} {480, 339}

\bibitem[\protect\citeauthoryear{{Raiteri} et~al.,}{{Raiteri}
  et~al.}{2017}]{2017Natur.552..374R}
{Raiteri} C.~M.,  et~al., 2017, \mn@doi [\nat] {10.1038/nature24623}, \href
  {https://ui.adsabs.harvard.edu/abs/2017Natur.552..374R} {552, 374}

\bibitem[\protect\citeauthoryear{{Rani}, {Wiita}  \& {Gupta}}{{Rani}
  et~al.}{2009}]{2009ApJ...696.2170R}
{Rani} B.,  {Wiita} P.~J.,   {Gupta} A.~C.,  2009, \mn@doi [\apj]
  {10.1088/0004-637X/696/2/2170}, \href
  {https://ui.adsabs.harvard.edu/abs/2009ApJ...696.2170R} {696, 2170}

\bibitem[\protect\citeauthoryear{{Remillard} \& {McClintock}}{{Remillard} \&
  {McClintock}}{2006}]{2006ARA&A..44...49R}
{Remillard} R.~A.,  {McClintock} J.~E.,  2006, \mn@doi [\araa]
  {10.1146/annurev.astro.44.051905.092532}, \href
  {https://ui.adsabs.harvard.edu/abs/2006ARA&A..44...49R} {44, 49}

\bibitem[\protect\citeauthoryear{{Romero}, {Cellone}  \& {Combi}}{{Romero}
  et~al.}{2000}]{RCC00aa}
{Romero} G.~E.,  {Cellone} S.~A.,   {Combi} J.~A.,  2000, \aap, 360, L47

\bibitem[\protect\citeauthoryear{{Romero}, {Cellone}, {Combi}  \&
  {Andruchow}}{{Romero} et~al.}{2002}]{RCCA02}
{Romero} G.~E.,  {Cellone} S.~A.,  {Combi} J.~A.,   {Andruchow} I.,  2002,
  \aap, 390, 431

\bibitem[\protect\citeauthoryear{{Roy}, {Sarkar}, {Chatterjee}, {Gupta},
  {Chitnis}  \& {Wiita}}{{Roy} et~al.}{2022}]{2022MNRAS.510.3641R}
{Roy} A.,  {Sarkar} A.,  {Chatterjee} A.,  {Gupta} A.~C.,  {Chitnis} V.,
  {Wiita} P.~J.,  2022, \mn@doi [\mnras] {10.1093/mnras/stab3701}, \href
  {https://ui.adsabs.harvard.edu/abs/2022MNRAS.510.3641R} {510, 3641}

\bibitem[\protect\citeauthoryear{{Sandrinelli}, {Covino}  \&
  {Treves}}{{Sandrinelli} et~al.}{2016}]{2016ApJ...820...20S}
{Sandrinelli} A.,  {Covino} S.,   {Treves} A.,  2016, \mn@doi [\apj]
  {10.3847/0004-637X/820/1/20}, \href
  {https://ui.adsabs.harvard.edu/abs/2016ApJ...820...20S} {820, 20}

\bibitem[\protect\citeauthoryear{{Sarkar}, {Kushwaha}, {Gupta}, {Chitnis}  \&
  {Wiita}}{{Sarkar} et~al.}{2020}]{2020A&A...642A.129S}
{Sarkar} A.,  {Kushwaha} P.,  {Gupta} A.~C.,  {Chitnis} V.~R.,   {Wiita} P.~J.,
   2020, \mn@doi [\aap] {10.1051/0004-6361/202038052}, \href
  {https://ui.adsabs.harvard.edu/abs/2020A&A...642A.129S} {642, A129}

\bibitem[\protect\citeauthoryear{{Sarkar}, {Gupta}, {Chitnis}  \&
  {Wiita}}{{Sarkar} et~al.}{2021}]{2021MNRAS.501...50S}
{Sarkar} A.,  {Gupta} A.~C.,  {Chitnis} V.~R.,   {Wiita} P.~J.,  2021, \mn@doi
  [\mnras] {10.1093/mnras/staa3211}, \href
  {https://ui.adsabs.harvard.edu/abs/2021MNRAS.501...50S} {501, 50}

\bibitem[\protect\citeauthoryear{{Scargle}}{{Scargle}}{1982}]{1982ApJ...263..835S}
{Scargle} J.~D.,  1982, \mn@doi [\apj] {10.1086/160554}, \href
  {https://ui.adsabs.harvard.edu/abs/1982ApJ...263..835S} {263, 835}

\bibitem[\protect\citeauthoryear{Schulz \& Mudelsee}{Schulz \&
  Mudelsee}{2002}]{SCHULZ2002421}
Schulz M.,  Mudelsee M.,  2002, \mn@doi [Computers & Geosciences]
  {https://doi.org/10.1016/S0098-3004(01)00044-9}, 28, 421

\bibitem[\protect\citeauthoryear{{Sillanp\"{a}\"{a}}, {Haarala}, {Valtonen},
  {Sundelius}  \& {Byrd}}{{Sillanp\"{a}\"{a}}
  et~al.}{1988}]{1988ApJ...325..628S}
{Sillanp\"{a}\"{a}} A.,  {Haarala} S.,  {Valtonen} M.~J.,  {Sundelius} B.,
  {Byrd} G.~G.,  1988, \mn@doi [\apj] {10.1086/166033}, \href
  {https://ui.adsabs.harvard.edu/abs/1988ApJ...325..628S} {325, 628}

\bibitem[\protect\citeauthoryear{{Smith}, {Balonek}, {Heckert}, {Elston}  \&
  {Schmidt}}{{Smith} et~al.}{1985}]{1985AJ.....90.1184S}
{Smith} P.~S.,  {Balonek} T.~J.,  {Heckert} P.~A.,  {Elston} R.,   {Schmidt}
  G.~D.,  1985, \mn@doi [\aj] {10.1086/113824}, \href
  {https://ui.adsabs.harvard.edu/abs/1985AJ.....90.1184S} {90, 1184}

\bibitem[\protect\citeauthoryear{{Smith}, {Montiel}, {Rightley}, {Turner},
  {Schmidt}  \& {Jannuzi}}{{Smith} et~al.}{2009}]{2009arXiv0912.3621S}
{Smith} P.~S.,  {Montiel} E.,  {Rightley} S.,  {Turner} J.,  {Schmidt} G.~D.,
  {Jannuzi} B.~T.,  2009, arXiv e-prints, \href
  {https://ui.adsabs.harvard.edu/abs/2009arXiv0912.3621S} {p. arXiv:0912.3621}

\bibitem[\protect\citeauthoryear{{Spinrad} \& {Smith}}{{Spinrad} \&
  {Smith}}{1975}]{1975ApJ...201..275S}
{Spinrad} H.,  {Smith} H.~E.,  1975, \mn@doi [\apj] {10.1086/153883}, \href
  {https://ui.adsabs.harvard.edu/abs/1975ApJ...201..275S} {201, 275}

\bibitem[\protect\citeauthoryear{{Stickel}, {Fried}  \& {Kuehr}}{{Stickel}
  et~al.}{1988}]{1988A&A...198L..13S}
{Stickel} M.,  {Fried} J.~W.,   {Kuehr} H.,  1988, \aap, \href
  {https://ui.adsabs.harvard.edu/abs/1988A&A...198L..13S} {198, L13}

\bibitem[\protect\citeauthoryear{{Sundelius}, {Wahde}, {Lehto}  \&
  {Valtonen}}{{Sundelius} et~al.}{1997}]{1997ApJ...484..180S}
{Sundelius} B.,  {Wahde} M.,  {Lehto} H.~J.,   {Valtonen} M.~J.,  1997, \mn@doi
  [\apj] {10.1086/304331}, \href
  {https://ui.adsabs.harvard.edu/abs/1997ApJ...484..180S} {484, 180}

\bibitem[\protect\citeauthoryear{{Takalo} et~al.,}{{Takalo}
  et~al.}{1998}]{1998A&AS..129..577T}
{Takalo} L.~O.,  et~al., 1998, \mn@doi [\aaps] {10.1051/aas:1998205}, \href
  {https://ui.adsabs.harvard.edu/abs/1998A&AS..129..577T} {129, 577}

\bibitem[\protect\citeauthoryear{{Tripathi}, {Gupta}, {Aller}, {Wiita},
  {Bambi}, {Aller}  \& {Gu}}{{Tripathi} et~al.}{2021}]{2021MNRAS.501.5997T}
{Tripathi} A.,  {Gupta} A.~C.,  {Aller} M.~F.,  {Wiita} P.~J.,  {Bambi} C.,
  {Aller} H.,   {Gu} M.,  2021, \mn@doi [\mnras] {10.1093/mnras/stab058}, \href
  {https://ui.adsabs.harvard.edu/abs/2021MNRAS.501.5997T} {501, 5997}

\bibitem[\protect\citeauthoryear{{Urry} \& {Padovani}}{{Urry} \&
  {Padovani}}{1995}]{1995PASP..107..803U}
{Urry} C.~M.,  {Padovani} P.,  1995, \mn@doi [\pasp] {10.1086/133630}, \href
  {https://ui.adsabs.harvard.edu/abs/1995PASP..107..803U} {107, 803}

\bibitem[\protect\citeauthoryear{{Valtonen} et~al.,}{{Valtonen}
  et~al.}{2008}]{2008Natur.452..851V}
{Valtonen} M.~J.,  et~al., 2008, \mn@doi [\nat] {10.1038/nature06896}, \href
  {https://ui.adsabs.harvard.edu/abs/2008Natur.452..851V} {452, 851}

\bibitem[\protect\citeauthoryear{{Valtonen} et~al.,}{{Valtonen}
  et~al.}{2019}]{2019ApJ...882...88V}
{Valtonen} M.~J.,  et~al., 2019, \mn@doi [\apj] {10.3847/1538-4357/ab3573},
  \href {https://ui.adsabs.harvard.edu/abs/2019ApJ...882...88V} {882, 88}

\bibitem[\protect\citeauthoryear{{Valtonen} et~al.,}{{Valtonen}
  et~al.}{2021}]{2021Galax..10....1V}
{Valtonen} M.~J.,  et~al., 2021, \mn@doi [Galaxies] {10.3390/galaxies10010001},
  \href {https://ui.adsabs.harvard.edu/abs/2021Galax..10....1V} {10, 1}

\bibitem[\protect\citeauthoryear{{Villata}, {Raiteri}, {Sillanp\"{a}\"{a}}  \&
  {Takalo}}{{Villata} et~al.}{1998}]{1998MNRAS.293L..13V}
{Villata} M.,  {Raiteri} C.~M.,  {Sillanp\"{a}\"{a}} A.,   {Takalo} L.~O.,
  1998, \mn@doi [\mnras] {10.1046/j.1365-8711.1998.01244.x}, \href
  {https://ui.adsabs.harvard.edu/abs/1998MNRAS.293L..13V} {293, L13}

\bibitem[\protect\citeauthoryear{{Villata} et~al.,}{{Villata}
  et~al.}{2002}]{2002A&A...390..407V}
{Villata} M.,  et~al., 2002, \mn@doi [\aap] {10.1051/0004-6361:20020662}, \href
  {https://ui.adsabs.harvard.edu/abs/2002A&A...390..407V} {390, 407}

\bibitem[\protect\citeauthoryear{{Villata} et~al.,}{{Villata}
  et~al.}{2008}]{2008A&A...481L..79V}
{Villata} M.,  et~al., 2008, \mn@doi [\aap] {10.1051/0004-6361:200809552},
  \href {https://ui.adsabs.harvard.edu/abs/2008A&A...481L..79V} {481, L79}

\bibitem[\protect\citeauthoryear{{Villata} et~al.,}{{Villata}
  et~al.}{2009}]{2009A&A...504L...9V}
{Villata} M.,  et~al., 2009, \mn@doi [\aap] {10.1051/0004-6361/200912732},
  \href {https://ui.adsabs.harvard.edu/abs/2009A&A...504L...9V} {504, L9}

\bibitem[\protect\citeauthoryear{{Wang} \& {Jiang}}{{Wang} \&
  {Jiang}}{2020}]{2020ApJ...902...41W}
{Wang} Y.-F.,  {Jiang} Y.-G.,  2020, \mn@doi [\apj] {10.3847/1538-4357/abb36c},
  \href {https://ui.adsabs.harvard.edu/abs/2020ApJ...902...41W} {902, 41}

\bibitem[\protect\citeauthoryear{{Zechmeister} \& {K{\"u}rster}}{{Zechmeister}
  \& {K{\"u}rster}}{2009}]{2009A&A...496..577Z}
{Zechmeister} M.,  {K{\"u}rster} M.,  2009, \mn@doi [\aap]
  {10.1051/0004-6361:200811296}, \href
  {https://ui.adsabs.harvard.edu/abs/2009A&A...496..577Z} {496, 577}

\bibitem[\protect\citeauthoryear{{Zhang}, {Yan}, {Liao}  \& {Wang}}{{Zhang}
  et~al.}{2017a}]{2017ApJ...835..260Z}
{Zhang} P.-f.,  {Yan} D.-h.,  {Liao} N.-h.,   {Wang} J.-c.,  2017a, \mn@doi
  [\apj] {10.3847/1538-4357/835/2/260}, \href
  {https://ui.adsabs.harvard.edu/abs/2017ApJ...835..260Z} {835, 260}

\bibitem[\protect\citeauthoryear{{Zhang}, {Zhang}, {Yan}, {Fan}  \&
  {Liu}}{{Zhang} et~al.}{2017b}]{2017ApJ...849....9Z}
{Zhang} P.,  {Zhang} P.-f.,  {Yan} J.-z.,  {Fan} Y.-z.,   {Liu} Q.-z.,  2017b,
  \mn@doi [\apj] {10.3847/1538-4357/aa8d6e}, \href
  {https://ui.adsabs.harvard.edu/abs/2017ApJ...849....9Z} {849, 9}

\bibitem[\protect\citeauthoryear{{Zhang}, {Zhang}, {Liao}, {Yan}, {Fan}  \&
  {Liu}}{{Zhang} et~al.}{2018}]{2018ApJ...853..193Z}
{Zhang} P.-f.,  {Zhang} P.,  {Liao} N.-h.,  {Yan} J.-z.,  {Fan} Y.-z.,   {Liu}
  Q.-z.,  2018, \mn@doi [\apj] {10.3847/1538-4357/aaa29a}, \href
  {https://ui.adsabs.harvard.edu/abs/2018ApJ...853..193Z} {853, 193}

\bibitem[\protect\citeauthoryear{{Zhou}, {Wang}, {Chen}, {Wiita},
  {Vadakkumthani}, {Morrell}, {Zhang}  \& {Zhang}}{{Zhou}
  et~al.}{2018}]{2018NatCo...9.4599Z}
{Zhou} J.,  {Wang} Z.,  {Chen} L.,  {Wiita} P.~J.,  {Vadakkumthani} J.,
  {Morrell} N.,  {Zhang} P.,   {Zhang} J.,  2018, \mn@doi [Nature
  Communications] {10.1038/s41467-018-07103-2}, \href
  {https://ui.adsabs.harvard.edu/abs/2018NatCo...9.4599Z} {9, 4599}

\makeatother
\end{thebibliography}








\bsp	
\label{lastpage}
\end{document}